\documentclass[letterpaper]{article} 
\usepackage[]{aaai25}  
\usepackage{times}  
\usepackage{helvet}  
\usepackage{courier}  
\usepackage[hyphens]{url}  
\usepackage{graphicx} 
\usepackage{natbib}  
\usepackage{caption} 
\usepackage{algorithm}
\usepackage{algorithmic}

\usepackage{newfloat}
\usepackage{listings}
\DeclareCaptionStyle{ruled}{labelfont=normalfont,labelsep=colon,strut=off} 
\floatstyle{ruled}
\newfloat{listing}{tb}{lst}{}
\floatname{listing}{Listing}
\title{AI Toolkit: Libraries and Essays for Exploring the Technology and Ethics of AI}
\author{
    Levin Ho\textsuperscript{\rm 1},
    Morgan McErlean\textsuperscript{\rm 1},
    Zehua You\textsuperscript{\rm 1},
    Douglas Blank\textsuperscript{\rm 2}, and
    Lisa Meeden\textsuperscript{\rm 1}
    }
\affiliations{
    \textsuperscript{\rm 1}Computer Science Department, Swarthmore College\\ 
    \textsuperscript{\rm 2}Comet ML, Head of Research\\
    \{lho2, mmcerle1, zyou1, lmeeden1\}@swarthmore.edu, doug.blank@gmail.com\\
}

\begin{document}

\maketitle


\begin{abstract}

In this paper we describe the development and evaluation of AITK, the Artificial Intelligence Toolkit. This open-source project contains both Python libraries and computational essays (Jupyter notebooks) that together are designed to allow a diverse audience with little or no background in AI to interact with a variety of AI tools, exploring in more depth how they function, visualizing their outcomes, and gaining a better understanding of their ethical implications. These notebooks have been piloted at multiple institutions in a variety of humanities courses centered on the theme of responsible AI. In addition, we conducted usability testing of AITK. Our pilot studies and usability testing results indicate that AITK is easy to navigate and effective at helping users gain a better understanding of AI. Our goal, in this time of rapid innovations in AI, is for AITK to provide an accessible resource for faculty from any discipline looking to incorporate AI topics into their courses and for anyone eager to learn more about AI on their own.  
\end{abstract}

\section{Introduction}

AI's impact on society is at an all time high. The One Hundred Year Study on AI states that ``the field’s successes have led to an inflection point: it is now urgent to think seriously about the downsides and risks that the broad application of AI is revealing'' \cite[p. 71]{ai100}. Furthermore, the report argues that the AI research community has a crucial role to play, sharing ``important trends and findings with the public in informative and actionable ways, free of hype and clear about the dangers and unintended consequences along with the opportunities and  benefits'' \cite[p. 71]{ai100}. 

Currently, a number of excellent resources exist to help explain AI, including blogs such as Janelle Shane's AI Weirdness \cite{shane} and YouTube channels such as Grant Sanderson's 3Blue1Brown \cite{sanderson}. Also there are numerous toolkits to help build AI such as the Open-Source AI Cookbook \cite{aicookbook}, Fast AI \cite{fastai} and Machine Learning Complete \cite{mlcomplete}. What is novel about the AITK project with respect to these existing resources and tools? Let's begin by describing AITK. 

AITK consists of both a set of Python libraries and a set of Jupyter notebooks. The Python libraries are designed to make it easy to experiment with key AI frameworks such as deep learning and robotics. The code focuses on allowing users to visualize the inner workings of these frameworks, opening up the black box. For example, when building a neural network, users can display the network's structure and visualize the network's activations across all layers based on any desired input values. But how can novices utilize these tools without knowing how to program? The Jupyter notebooks provide a way for novices to interact with this code without having to write it themselves. A notebook consists of a series of cells, where each cell contains either executable code or formatted text. These notebooks enable users to interactively execute and modify prewritten code, interspersed with accompanying explanation, allowing them to gain a richer understanding of AI and how it functions. One can think of Jupyter notebooks as essentially computational essays, and they have transformed the way science is now communicated \cite{nature}.

Our aim with the AITK project is to provide a resource for better understanding AI and its risks that is free, open-source, accessible to novices, and fundamentally interactive. Blogs and YouTube channels provide informative analysis of AI, but they lack the ability for users to try out the concepts on their own. 
Existing educational toolkits, including those using Jupyter notebooks, are generally about teaching users to build their own AI and so require substantial coding ability. As a result, they are not well-suited for novices, or for those mainly wishing to become better informed about AI. The AITK project bridges the gap between explanations and experimentation, by providing both together as computational essays that were written with novices in mind. 

\begin{figure*}[h]
    \centering
    \includegraphics[width=0.9\linewidth]{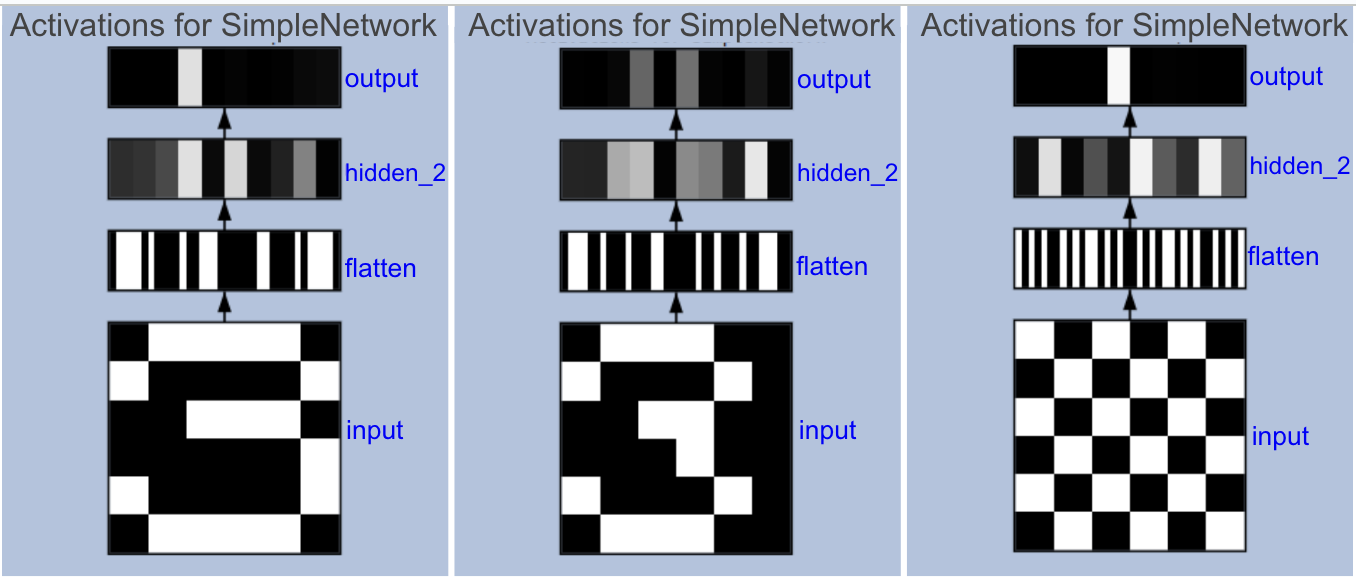}
    \caption{AITK allows users to visualize the activations within a network. On the left, the network has correctly recognized the image as the digit 3. In the center, the network is unsure of whether the image represents the digit 3 or 5. On the right, the network has incorrectly classified the checkerboard image as the digit 4.}
    \label{network}
\end{figure*}

The rest of this paper is organized as follows. First, we describe the types of computational essays that are available in AITK, and provide details about one example notebook. Next, we discuss how AITK has already been piloted at a number of different institutions in humanities courses focused on the theme of responsible AI. Then we describe how AITK can help address the new focus of Computer Science curricula on the societal impacts of AI. Because our goal is that AITK should be accessible to anyone, we also conducted usability testing and describe both the design and results of this testing. Based on the results, we reflect on ways to improve AITK. Finally, we summarize the paper in the Conclusions. 

\section{Description of AITK}

AITK began as a collaborative effort between three computer science faculty members, Douglas Blank, James Marshall, and Lisa Meeden, built on many years of experience teaching AI at small liberal arts colleges, and based on several previously created open source projects \cite{pyro2006,ohara}. It was released as an open source project on GitHub\footnote{https://github.com/ArtificialIntelligenceToolkit/aitk} in 2021. It underwent significant additions and upgrades in 2024 with the help of three undergraduate students (all co-authors on this paper).

\begin{table*}[ht]
    \centering
    \begin{tabular}{l|l|l|l}
              &  {\bf Neural Networks} & {\bf Generative AI} & {\bf Robotics} \\ \hline
     {\bf Begin}    &  {\it Basic Neural Networks} & {\it Word Embedding} & {\it What is it like to be a robot?} \\ \hline
     {\bf Next} & {\it Categorizing Faces} & {\it Nano GPT} & {\it Braitenberg Vehicles} \\
          & {\it Data Manipulation}  & {\it Image Generation} & {\it Seek Light} \\
\hline
    {\bf Further} & {\it Analyzing Hidden Representations} & {\it Transformer} & {\it Subsumption} \\
                  & {\it Structure of Convolutional Networks} & & \\
     
    \end{tabular}
    \caption{Sequencing of AITK notebooks by category}
    \label{sequencing}
\end{table*}

The notebooks within AITK are organized into three main categories: Neural Networks, Generative AI, and Robotics. There is also an Advanced category that is most appropriate for users with a deeper background in computer science and AI. A suggested sequencing is provided for where to begin and how to proceed within each category, see Table \ref{sequencing}.

\begin{figure}[h]
    \centering
    \includegraphics[width=0.9\linewidth]{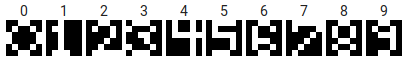}
    \caption{A sample of some of the 6x6 digit images used in the {\it Basic Neural Networks} notebook.}
    \label{digits}
\end{figure}

\subsection{Example: {\it Basic Neural Networks} Notebook}

Because neural networks and deep learning are the foundation of much of the latest innovations in AI, we recommend beginning with the Neural Networks category, and specifically with the notebook {\it Basic Neural Networks}, which focuses on a small, simple dataset of 6x6 digits, see Figure~\ref{digits}. The task in this notebook is to train a neural network to correctly classify the images as representing the digits 0-9. The notebook introduces fundamental concepts from machine learning such as categorization, training and validation sets, generalization, and overfitting. 

\begin{figure}
    \centering
    \includegraphics[width=1\linewidth]{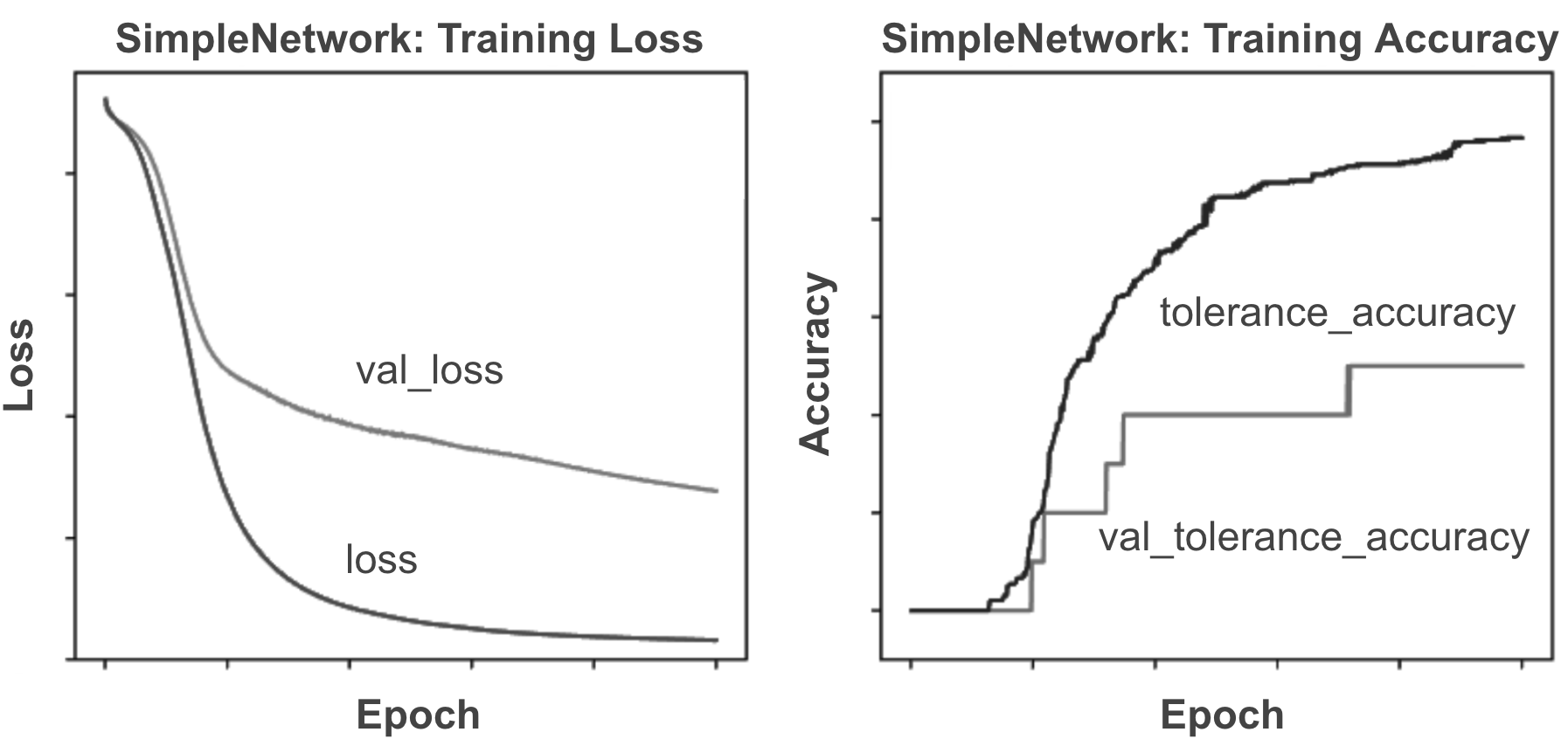}
    \caption{AITK automatically generates graphs summarizing the network's progress during training at reducing loss (at left) and improving accuracy (at right).}
    \label{training}
\end{figure}

One of the key features of AITK is that it provides numerous visualizations to make the the concepts being introduced more concrete. For example, AITK allows the user to view a depiction of the network. The sequence of layers in the network are shown as boxes connected by arrows, and activation values within each layer are encoded in grayscale, with black representing 0, white representing 1, and gray for values in between, see Figure~\ref{network}. Users can see from this visualization that the network has a two-dimensional input layer, which is flattened, and passed through one hidden layer to the output layer. During training, AITK also produces graphs showing the network's progress at both reducing loss and improving accuracy, see Figure~\ref{training}.

\begin{figure}[h]
    \centering
    \includegraphics[width=0.9\linewidth]{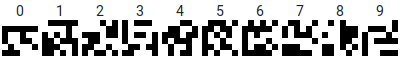}
    \caption{A sample of 6x6 random images generated in the {\it Basic Neural Networks} notebook for training on what is {\bf not} a digit.}
    \label{random}
\end{figure}

Users are invited to draw their own 6x6 digit and test whether the trained network is able to correctly categorize it, demonstrating the idea of generalization to novel data. In addition, the notebook explores how the network responds to other patterns, such as a checkerboard, and shows that the network may ``recognize'' it as a particular digit (see Figure~\ref{training}, right), which is clearly problematic. This leads to a second experiment where random images are added to the training set (replacing all of the images of the digit zero). We then test if we can train the network to recognize both when an image is {\bf not} a digit (see Figure~\ref{random} for examples of images that are not digits) and when an image is one of the digits one through nine. Although, this experiment is only partially successful, the second version of the network is typically able to learn that the checkerboard is {\bf not} a digit. 

Before using this notebook, many users have incorrect intuitions about how neural networks operate. They often expect the network to correctly classify every image, never making mistakes (as shown in the center and right of Figure~\ref{network}). They are often surprised that each training run of the network may yield different outcomes, due to the random initialization of the weights. And by changing the data set and the task in the second experiment, they begin to recognize the importance of the training data in shaping the outcome (this theme is explored in greater depth in the notebook {\it Data Manipulation}). As AI is used in more and more critical applications, gaining these insights plays an important role in better informing the public about how AI works and its potential dangers. 

Next we will explore how AITK notebooks, such as the one discussed in this section, have been piloted in a variety of college courses. 

\section{Piloted in Humanities Courses}

Starting in 2021, the National Humanities Center brought together faculty from fifteen institutions to develop courses to engage students on their respective campuses to think more deeply about the ethical issues surrounding AI. This initiative was called the {\it Responsible Artificial Intelligence Curriculum Design Project} \cite{nhc}. The cohort of faculty spent two years developing new humanities courses and offered them for the first time during the 2023-24 academic year. 

Notebooks from AITK were piloted in a subset of these courses: those developed at Bowdoin College, Davidson College, Duke University, Swarthmore College, and the University of Utah. All of these courses used the {\it Basic Neural Networks} notebook described earlier. Each faculty member downloaded a copy of the AITK notebook to Google Colab, making it easy for them to customize the notebook for their needs, elaborating on or removing the existing material. Then students made their own copies of the faculty member's notebook and also ran them in Google Colab. The requirements to using one of these notebooks in a course is relatively lightweight: students must have a Google account and a laptop with a web browser. During these pilot studies all students were able to successfully use and complete the notebooks. 

\subsection{Case Study: Swarthmore College}

Let's look more closely at one of these course entitled {\it Ethics and Technology}, co-developed and co-taught by a Computer Scientist and a Philosopher ~\cite{course} at Swarthmore College. The prerequisite for this course was an introductory course in either Computer Science or Philosophy, leading to a diverse mix of student backgrounds. Students were expected to write philosophy papers, and to complete labs related to AI, where the labs were based largely on AITK notebooks. Initially, the CS students were nervous about the papers, while the Philosophy students were worried about the labs. 

This course began with primers on both ethics and AI. The notebook {\it Basic Neural Networks} was used during the AI primer. Throughout the semester, the students read chapters from Kate Crawford's book {\it Atlas of AI}~\cite{crawford}. The notebook {\it Categorizing Faces} was used to illustrate concepts from her chapters entitled ``Classification'' and ``Affect''. Other readings were also included to delve into additional topics not covered by Crawford. For example, we discussed embodiment and read Thomas Nagel's well-known paper {\it What is it like to be a bat?}~\cite{nagel} and paired this with the AITK notebook {\it What is it like to be a robot?}. 

In the final course evaluations, students were asked ``How did the lab component of the course complement your understanding of the course's content?'' Here is a sample of their responses:

\begin{itemize}
    \item ``The lab component was very helpful. It was quite lightweight (people without a lot of experience will understand).''
    \item ``The lab was super helpful to put our arguments in class into context.''
    \item ``I enjoyed the hands-on nature of the labs and the supplemental questions helped with my understanding.''
\end{itemize}

Some of the students with CS background wished the labs had been more substantial:

\begin{itemize}
     \item ``Loved the labs---maybe more CS intensity can be introduced. It was a fun way to gain insights.''  
     \item ``As a CS major they were quite easy, that being said they served a purpose.''
     \item ``I loved the labs! I wish they had gotten a little more technical, had us create neural networks with slightly less hand holding, but I feel they gave me a really good understanding of how ML works.''
\end{itemize}

We were encouraged that the humanities majors in this course found the AITK notebooks to be accessible and that the CS majors were still engaged despite wanting to do more technical work. 

After the successful use of AITK notebooks in this undergraduate course in spring 2024, we piloted them a second time in a four-week mini-course on AI in summer 2024 for faculty and staff at Swarthmore College, which was sponsored by the Teaching and Learning Commons. There was a diverse mix of participants in this mini-course including faculty and lecturers from departments in the Humanities, Social Sciences, and Sciences, as well as staff from the Library, Admissions, Student Affairs, Advancement, Technology Services, and Career Services. At the end of this course, we asked the participants to reflect on the use of the AITK notebooks for lab work. Here are a few of their responses:

\begin{itemize}
    \item ``The notebooks were terrific, going back later to read everything and run the code again really reinforced the topics.''
    \item ``I was really intimidated by the idea of doing labs, but in reality they weren't scary at all. I like that I am able to save and refer back to the notebooks.''
\end{itemize}

In this section, we've seen how AITK notebooks have been successfully used in both humanities courses at multiple institutions and for a more general audience in a summer course on AI. Next let's consider how AITK notebooks could address new emerging requirements for CS undergraduate curricula to focus on the societal impacts of AI.

\begin{figure*}[h]
\centering
\includegraphics[width=0.8\textwidth]{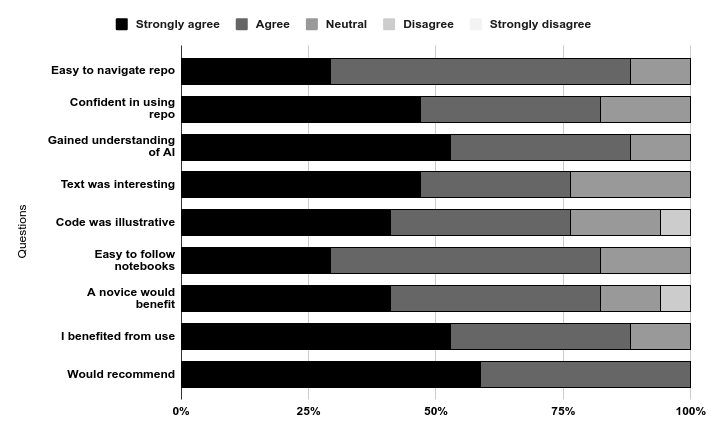}
\caption{User testing responses indicate that AITK is easy to use and accessible for most novices.}
\label{users}
\end{figure*}
 
\section{Highlighting Societal Impacts of AI}

Approximately every 10 years the ACM and IEEE, the primary professional organizations of computer science, issue updated curricular recommendations for undergraduate CS education. The most recent update, called CS2023, ``highlights how important it is to understand and assess the broader societal impacts and implications of AI methods and applications, including issues in AI ethics, fairness, trust, and explainability'' ~\cite{acm2023}. To reflect this, a new core knowledge unit has been added entitled ``Applications and Societal Impact'' with a suggested 3 hours of instruction. 

We believe that a number of AITK notebooks offer an excellent resource for helping to meet these requirements. For example the following notebooks focus on the societal impacts of AI:

\begin{description}
    \item[{\it Data Manipulation}] Examines how the relative numbers within each class of a data set biases the outcomes. It highlights the ``Gender Shades'' work that revealed biases in facial recognition software ~\cite{gendershades}.
    \item[{\it Word Embedding}] Describes how word embeddings are formed and discusses how they learn biases implicit in training texts such as the association of particular careers with a specific gender ~\cite{wordembedding}.
    \item[{\it Image Generation}] Explains the process of image generation and concludes by revealing the biases that emerge in generated images, for example in typical skin tones produced for the prompt ``lawyer'' vs the prompt ``criminal'' ~\cite{bloomberg}.
\end{description}

Given the diverse audiences from both the humanities and the sciences that might be interested in using AITK, we decided to conduct a usability study. 

\section{Usability Study}

We had two goals in conducting the study. Our first goal was to observe how new users to AITK were able to navigate the GitHub website in order to find a notebook of interest to them. To this end, our user testing facilitators were trained by our UX Librarian who provided various materials to prepare them, such as \cite{uxtesting}. Our second goal was to assess the value of our notebooks as educational tools.

We recruited a total of 17 participants who were divided into three groups based on their experience level with CS and GitHub. We had 6 novice users who had never used GitHub and only one of whom had taken an introductory CS course. We had 6 intermediate users who had used GitHub and had taken at least one intermediate CS course. We had 5 experienced users, three were CS or math majors with several upper-level CS courses or some other relevant AI/ML experience and two were CS faculty members. 

The usability test lasted for at most one hour and participants received a \$25 gift card for their time. Their task was described by the facilitator as follows:

\begin{quote}
    We'll be doing a website usability session: I'll describe some situations for which information from the toolkit will be helpful and you’ll explore the toolkit to look for it. We'll be testing the toolkit---not you--in order to see if the toolkit is intuitive to use. It will help us to watch you navigate, especially if you are not able to find information, because that will tell us how to make the toolkit easier to use. Here is a topic to explore. Go through any notebooks necessary to gain knowledge in this topic.
\end{quote}

The landing page of the AITK website contains a link to a short summary of the notebook topics and another link to our suggested sequencing through the notebooks. Some participants read the summary and sequencing information first while others chose to click directly on the notebooks folder and began searching around for a notebook with their assigned topic. About half of the participants were able to successfully navigate to their desired notebook on the first try, while the other half needed some guidance.  

Once a participant reached the notebook of interest, they were asked to use the notebook, executing the cells within it. For participants who had never used a notebook before, there were some challenges. Several wondered whether they needed to actually read through the code and understand it (this wasn't our intent). In some cases, participants forgot to execute a code block, and then later code blocks failed due to missing dependencies, and they didn't understand how to fix it based on the error messages. 

In contrast, during our pilot studies in courses, students had been guided through obtaining the notebook they needed and help was on hand when using the notebooks, so these issues had not arisen. It was useful to discover that for novices working on their own, both navigating the website and using the notebooks could be challenging. 

At the end of the usability testing, participants were asked to complete a survey. They rated their agreement with a series of statements using a 5-point Likert scale: Strongly agree, Agree, Neutral, Disagree, or Strongly Disagree. The statements they were surveyed about were:

\begin{itemize}
    \item The GitHub repo was easy to follow and the sequencing of notebooks was logical to me.
    \item I am confident in my ability to navigate this GitHub repo and use the notebooks on my own.
    \item I feel more confident in my understanding of the topic I reviewed.
    \item I found the text sections of the notebook interesting.
    \item I found the code sections (running the code cells, visualizations, etc.) engaging and illustrative.
    \item I found the notebook accessible and easy to follow.
    \item I believe a novice (a person who doesn’t have a lot of background knowledge) would benefit from using this tool.
    \item I believe that my time using this toolkit has benefited me.
    \item Based on the content I saw, I would recommend this toolkit to others.
\end{itemize}

A summary of the user testing results are provided in Figure~\ref{users}. For every statement at least 75\% of the participants strongly agreed or agreed with them. 

For two of the statements, about the code sections being illustrative and novices benefiting, a single participant disagreed in each case. Based on this result and on our observations during user testing, we plan to provide more scaffolding for novices as they use the notebooks. For example, we will periodically remind users that when an error occurs, it is likely due to forgetting to execute a previous code cell, and we will suggest that they go back and rerun cells from the beginning. We also plan to explain that users do not need to understand the code sections; they simply need to run them to see the results. 

We were heartened to see that the notebooks did succeed as an educational tool, with 87\% of the participants agreeing that they understood a topic better after using AITK and that they benefited from its use. Also, 100\% of the participants agreed that they would recommend AITK to others. 

\section{Conclusions}

In this paper we described the Artificial Intelligence Toolkit (known as AITK), an open-source project consisting of both Python libraries and Jupyter notebooks. Our goal for this project is to provide a resource for better understanding AI that is free, accessible to novices, and fundamentally interactive. We argued that though many other resources exists for exploring AI, they are either too advanced for novices, or are simply explanations, with no opportunity for interaction. AITK bridges the gap between experimentation and explanation by providing both components together within computational essays that were written with novices in mind. To illustrate this, we provided details on one notebook that introduces users to key concepts related to neural networks and machine learning. 

Next, we described how AITK has been successfully piloted in humanities courses developed around the theme of responsible AI at a number of institutions involved in a National Humanities Center initiative. The fact that humanities students, who for the most part had no computer science background, were able to successfully use these notebooks demonstrates that they are accessible to novices. We argued, that AITK could also help computer science faculty address the latest ACM/IEEE curriculum requirements for a new core knowledge unit that focuses on AI applications and their societal impacts. We noted that AITK is a lightweight addition to any course, only requiring students to have a Google account, and a laptop with a web browser to be able to run the notebooks within Google Colab. Finally, we conducted a usability study of AITK with a mix of novice, intermediate, and advanced participants. Results were overwhelmingly positive, though our observations revealed that there are still ways to improve the toolkit by providing additional scaffolding for novices. 

\section*{Acknowledgments}

This work was supported in part by a grant from the National Humanities Center: {\it Responsible Artificial Intelligence Curriculum Design Project} \cite{nhc}. 

\bibliography{SwatAAAI25}

\end{document}